\documentclass[12pt,preprint]{aastex}

\usepackage{apjfonts}
\def\Meszaros{M\'esz\'aros~}

\begin{document}


\title{High-energy gamma-ray afterglows from low-luminosity gamma-ray bursts}

\author{Hao-Ning He\altaffilmark{1}, Xiang-Yu Wang\altaffilmark{1,5}, Yun-Wei Yu\altaffilmark{1,2} and Peter M\'esz\'aros\altaffilmark{3,4}}
\altaffiltext{1}{Department of Astronomy, Nanjing University,
Nanjing 210093, China} \altaffiltext{2}{Institute of Astrophysics,
Huazhong Normal University, Wuhan 430079, China}
\altaffiltext{3}{Department of Astronomy and Astrophysics,
Pennsylvania State University, University Park, PA 16802, USA}
\altaffiltext{4}{Department of Physics, Pennsylvania State
University, University Park, PA 16802, USA} \altaffiltext{5}{Key
Laboratory for Particle Astrophysics, Institute of High Energy
Physics, Chinese Academy of Sciences,  Beijing 100049, China}

\begin{abstract}
{The observations of gamma-ray bursts (GRBs) such as 980425, 031203
and 060218, with luminosities much lower than those of other classic
bursts, lead to the definition of a new class of GRBs --
low-luminosity GRBs. The nature of the outflow responsible for them
is not clear yet. Two scenarios have been suggested: one is the
conventional relativistic outflow with initial Lorentz factor of
order of $\Gamma_0\ga 10$ and the other is a trans-relativistic
outflow with $\Gamma_0\simeq 1-2$. Here we compare the high energy
gamma-ray afterglow emission from these two different models, taking
into account both synchrotron self inverse-Compton scattering (SSC)
and the external inverse-Compton scattering due to photons from the
cooling supernova or hypernova envelope (SNIC). We find that the
conventional relativistic outflow model predicts a relatively high
gamma-ray flux from SSC at early times ($<10^4 {\rm s}$ for typical
parameters) with a rapidly decaying light curve, while in the
trans-relativistic outflow model, one would expect a much flatter
light curve of high-energy gamma-ray emission at early times, which
could be dominated by both the SSC emission and SNIC emission,
depending on the properties of the underlying supernova and the
shock parameter $\epsilon_e$ and $\epsilon_B$. The Fermi Gamma-ray
Space Telescope should be able to distinguish between the two models
in the future.}

\end{abstract}


\section{INTRODUCTION}\label{introduction}
Long duration gamma-ray bursts are generally believed to result from
the death of massive stars, and their association  with
core-collapse supernovae (SNe of Type Ib/c) has been observed over
the last decade. The first hint for such a connection came with the
discovery of a nearby SN 1998bw in the error circle of GRB 980425
(Galama et al. 1998; Iwamoto et al. 1998) at distance of only about
$40 {\rm Mpc}$. The isotropic gamma-ray energy release is  of the
order of only $10^{48} {\rm erg}$ (Galama et al. 1998) and the radio
afterglow modelling suggests an energy of $10^{49}-10^{50} {\rm
erg}$ in a mildly relativistic ejecta (Kulkarni et al. 1998;
Chevalier \& Li 1999). Recently {\it Swift} discovered GRB 060218,
which is the second nearest GRB identified so far (Campana et al.
2006; Cusumano et al. 2006; Mirabal \& Halpern 2006; Sakamoto 2006).
It is also an under-energetic burst with energy in prompt
$\gamma$/X-rays about $5\times10^{49} {\rm erg}$ and is associated
with SN 2006aj. { Another burst, GRB 031203, is a third example of
this group (Sazonov et al. 2004)}.

{If one assumes that GRB 060218-like bursts follow the logN-logP
relationship of high-luminosity GRBs then, as argued by Guetta et
al. (2004), no such burst with redshifts $z < 0.17$ should be
observed by a HETE-like instrument within the next 20 years.
Therefore, the unexpected discovery of GRB 060218 may suggest that
these objects form} a different new class of GRBs from the
conventional high-luminosity GRBs (Soderberg et al. 2006; Pian et
al. 2006; Cobb et al. 2006; Liang et al. 2006; Guetta \& Della
Valle 2007; Dai 2008), although what distinguishes such
low-luminosity GRBs (LL-GRBs) from the conventional
high-luminosity GRBs (HL-GRBs) remains unknown. Their rate of
occurrence may be one order of magnitude higher than that of the
typical ones (e.g. Soderberg et al. 2008). They are rarely
recorded because such { intrinsically dim GRBs can only be
detected from relatively short distances} with present gamma-ray
instruments.

The "compactness problem" of GRBs requires that the outflow of
normal GRBs should be highly relativistic with a bulk Lorentz factor
of $\Gamma_0\ga 100$ (Baring \& Harding 1997; Lithwick \& Sari
2001). The low-luminosity and softer spectra of low-luminosity GRBs
relax this constraint on the bulk Lorentz factor. It was suggested
that the softer spectrum and low energetics of GRB060218 (or
classified as X-ray flash 060218) may indicate a somewhat lower
Lorentz factor of the order of $\sim 10$ (e.g. Mazzali 2006; Fan et
al. 2006; Toma et al. 2006).

An alternative possibility is that low-luminosity GRBs are driven
by a trans-relativistic outflow with $\Gamma_0\simeq2$ (Waxman 2004;
Waxman, M\'esz\'aros \& Campana, 2007; Wang, Li, Waxman \&
M\'esz\'aros  2007; Ando \& M\'esz\'aros 2008). The flat light
curve of the X-ray afterglow of the nearest GRB, GRB 980425/SN
1998bw, up to 100 days after the burst, has been argued to
result from the coasting phase of a mildly relativistic shell
with an energy of a few times $10^{49}$ erg (Waxman 2004). From the
thermal energy density in the prompt emission of
GRB060218/SN2006aj, Campana et al. (2006) inferred that the shell
driving the radiation-dominated shock in GRB 060218/SN 2006aj must
be mildly relativistic. This trans-relativistic shock could be
driven by the outermost parts of the envelope that get accelerated
to a mildly relativistic velocity when the supernova shock
accelerates in the density gradient of the envelope of the
supernova progenitor (Colgate 1974; Matzner \& McKee 1999; Tan et
al. 2001), or it could be due to a choked relativistic jet propagating
through the progenitor (Wang et al. 2007).

In this paper, we investigate the high energy afterglow emission
from low luminosity GRBs for both {relativistic and
trans-relativistic} models and explore whether the Fermi Gamma-ray
Space Telescope can distinguish between these two models with future
observations of low-luminosity GRBs. Since photons from the
underlying supernova are an important seed photon source for
inverse-Compton (IC) scattering, we consider both synchrotron
self-inverse Compton (SSC)  and external IC scattering due to
supernova photons (denoted by SNIC hereafter). At early times
(within a few days after the burst), a UV-optical SN component was
recently detected from SN2006aj and SN2008D (Campana et al. 2006;
Soderberg et al. 2008), which has been interpreted as the cooling SN
envelope emission after being heated by the radiation-dominated
shock\footnote{Although there is a disagreement on the origin of the
early UV-optical emission from SN2006aj, an agreement has been
reached for that of SN2008D (Waxman et al. 2007; Soderberg et al.
2008; Chevalier \& Fransson 2008).} (Waxman et al. 2007; Soderberg
et al. 2008; Chevalier \& Fransson 2008). In our calculation, we
take into account this seed photon source in addition to the
late-time supernova emission, which peaks after ten days.

Recently, Ando \& Meszaros (2008) discussed the broadband emission
from SSC and SNIC for a trans-relativistic ejecta in a
low-luminosity GRB at a particular time-- the ejecta deceleration
time. Here we study the time evolution of the high-energy
gamma-ray emission resulted from such IC processes and consider
both the trans-relativistic ejecta and the highly relativistic
ejecta scenarios for low-luminosity GRBs.

The paper is organized as follows. First, we describe the dynamics
of shock evolution in the two models in $\S$ \ref{models}.  In
$\S$3, we present the formula for the calculation of the
inverse-Compton emissivity. For the SNIC emission, we take into
account the anisotropic scattering effect.  Then we present the
results of the spectra and light curves of SSC and SNIC emission
for the two different models and explore the detectability of
these components by Fermi Large Area Telescope (LAT) in $\S$ 4.
Finally, we give the conclusions and discussions.

\section{Dynamics and Electron Energy Distribution}\label{models}

{We assume that the two models have the same parameters except for
the initial Lorentz factor of the ejecta. For the latter, we adopt
nominal values of}  $\Gamma_0=10$  in the conventional
relativistic ejecta model and  $\Gamma_0=2$ in the
trans-relativistic ejecta model, respectively. Note that in the
conventional relativistic ejecta model, even if $\Gamma_0\gg 10$
the dynamics of the blast wave is identical to the case of
$\Gamma_0= 10$ from the time tens of seconds after the burst,
because the blast wave has entered the Blandford-McKee
self-similar phase since then.

We consider a spherical GRB ejecta carrying a total energy of
$E=10^{50}E_{50}\rm{erg}$ expanding into a surrounding wind medium
with density profile $n=Kr^{-2}$, where $K\equiv\dot{M}/(4\pi
m_pv_{\rm w})=3\times10^{35}{\rm cm}^{-1}\dot{m}$ with
$\dot{m}\equiv(\dot{M}/10^{-5}M_{\odot}\rm{yr^{-1}})/(v_{\rm
w}/10^3\rm{kms^{-1}})$. As the circumburst medium is swept up by
the blast wave, the total kinetic energy  of the fireball is
(Panaitescu et al. 1998)
\begin{equation}
E_{\rm k}=(\gamma-1)(m_{\rm ej}+m)c^{2}+\gamma U'\label{energy}
\end{equation}
where $\gamma$ is the bulk Lorentz factor of the shell, $m_{\rm ej}$
the ejecta mass, $m$ is the mass of the swept-up medium, and the
comoving internal energy $U'$ can be expressed by
$U'=(\gamma-1)mc^2$, which is suitable for both ultrarelativistic
and Newtonian shocks (Huang et al. 1999). Hereafter superscript
prime represents that the quantities are measured in the comoving
frame of the shell. As usual, we assume that the magnetic field and
the electrons have a fraction $\epsilon_B$ and $\epsilon_e$ of the
internal energy, respectively. Following Eq. (\ref{energy}), the
differential dynamic equation can be derived as (Huang et al. 2000)
\begin{equation}
\frac{d\gamma}{dm}=-\frac{\gamma^{2}-1}{m_{\rm ej}+2\gamma m},
\end{equation}
which describes the overall evolution of the shell from relativistic
phase to non-relativistic phase. The initial value of $\gamma$ is
$\Gamma_0=E/(m_{\rm ej}c^2)$. To obtain the time-dependence of
$\gamma$ one makes use of
\begin{equation}
\frac{dm}{dR}=4\pi R^{2}nm_{\rm p},
\end{equation}
\begin{equation}
\frac{dR}{dt}=\beta c\gamma(\gamma+\sqrt{\gamma^{2}-1}),
\end{equation}
where $t$ is the observer time, $R$ and
$\beta=\sqrt{1-\gamma^{-2}}$ are the radius and velocity of the
shell, respectively.

Solving Eqs. 2, 3, and 4, three dynamic phases can be found: (i)
Coasting phase. The shell does not decelerate significantly until it
arrives at the deceleration radius, $R_{\rm dec}=E/(4\pi
K\Gamma_{0}^{2}m_{\rm p}c^{2})$, where the mass of the swept-up
medium $m$ is comparable to $m_{\rm ej}/\Gamma_{0}$ (Sari  \& Piran
1995). The corresponding deceleration time can be calculated from
$t_{\rm dec}\simeq R_{\rm dec}/(2\Gamma_0^2c)$. For representative
parameter values $E_{50}=1$ and $\dot{m}=1$, $R_{\rm
dec}\simeq4.4\times10^{15}\rm{cm}$ and $t_{\rm
dec}\simeq1.8\times10^4\rm{s}$ for a trans-relativistic ejecta with
$\Gamma_0=2$,  while $R_{\rm dec}\simeq1.8\times10^{14}\rm{cm}$ and
$t_{\rm dec}\simeq29\rm{s}$ for a conventional highly relativistic
ejecta with $\Gamma_0=10$; (ii) Blandford \& McKee self-similar
phase ($t>t_{\rm dec}$ and $\gamma\ga 2 $), where $\gamma\propto
t^{-1/4}$ and $R\propto t^{1/2}$; (iii) Non-relativistic phase
($\gamma\rightarrow1$), where $\beta\propto t^{-1/3}$ and $R\propto
t^{2/3}$, i.e., the Sedov-von Neumann-Taylor solution applies
(Zel'dovich $\&$ Raizer 2002, p.93, Waxman, 2004).

In the absence of radiation losses, the energy distribution of
shock-accelerated electrons behind the shock is usually assumed to
be a power-law as $dN_{e}/d\gamma_{e}\propto\gamma_{e}^{-p}$. As
the electrons are cooled by synchrotron and IC radiation, the
electron energy distribution becomes a broken power-law, given by

(1) for $\gamma_{e,\rm c}\leq\gamma_{e,\rm m}$,
\begin{equation}
\frac{dN_{e}}{d\gamma_{e}}\propto\left\{\begin{array}{ll}
\gamma_{e}^{-2},&\gamma_{e,\rm c}\leq\gamma_{e}\leq\gamma_{e,\rm m}\\
\gamma_{e}^{-p-1},&\gamma_{e,\rm m}<\gamma_{e}\leq\gamma_{e,\rm
max}\end{array}\right.
\end{equation}

(2) for $\gamma_{e,\rm m}<\gamma_{e,\rm c}\leq\gamma_{e,\rm max}$,
\begin{equation}
\frac{dN_{e}}{d\gamma_{e}}\propto\left\{\begin{array}{ll}
\gamma_{e}^{-p},&\gamma_{e,\rm m}\leq\gamma_{e}\leq\gamma_{e,\rm c}\\
\gamma_{e}^{-p-1},&\gamma_{e,\rm c}<\gamma_{e}\leq\gamma_{e,\rm
max}\end{array}\right.
\end{equation}
which are normalized by the total number of the electrons solved
from the dynamic equations. The minimum, cooling, and maximum
Lorentz factors of electrons are, respectively, given by
\begin{eqnarray}
\gamma_{e,\rm m}&=&\epsilon_{e}\frac{p-2}{p-1}\frac{m_{\rm
p}}{m_{e}}(\gamma-1)=92f_{p1}\epsilon_{e,-0.5}(\gamma-1)
\\
\gamma_{e,\rm c}&=&\frac{6\pi m_{e}c}{(1+Y)\sigma_{T}{B'}^{2}
(\gamma+\sqrt{\gamma^{2}-1})t}\nonumber\\
&&\simeq\frac{1.7\times10^3R_{15}^2} {\epsilon_{B,-3}\dot{m} t_{4} Y
(\gamma+\sqrt{\gamma^2-1})\gamma (\gamma-1)}
\\
\gamma_{e,\rm max}&=&\sqrt{6\pi q_{e}\over\sigma_{T}B'(1+Y)}\\
&&\simeq4.5\times10^7R_{15}\gamma^{-1/4}(\gamma-1)^{-1/4}Y^{-1/2}
\end{eqnarray}
where $f_{p1}=6(p-2)/(p-1)$, $B'$ is the comoving magnetic field
strength and $Y$ is the Compton parameter that is defined as the
ratio of the IC luminosity (including SSC and SNIC) to the
synchrotron luminosity\footnote{At very late times, when
$\gamma_{e,\rm m}$ decreases to be close to  a few, $\gamma_{e,\rm
m}=\epsilon_{e}\frac{p-2}{p-1}\frac{m_{\rm p}}{m_{e}}(\gamma-1)+1$
is used (Huang \& Cheng 2003).}.

\section{Inverse-Compton Emission}
The accelerated electrons can be cooled by synchrotron radiation and
inverse Compton (IC) scattering of seed photons (including
synchrotron photons and blackbody photons emitted by the supernova).
Since the IC emissivity on the basis of the Thomson cross section is
inaccurate for high-energy $\gamma$-rays, we use the full
Klein-Nishina cross section instead. Once the electron distribution
and the flux of seed photons ($f'_{\nu'_{\rm s}}$) (the distribution
of which is isotropic) are known, the IC emissivity (at frequency
$\nu'$) of electrons can be calculated by(Blumenthal $\&$ Gould
1970; Yu et al. 2007)
\begin{equation}\label{SSCiso}
{\varepsilon'}^{\rm IC}_{\rm iso}(\nu')=3\sigma_{\rm
T}\int^{\gamma_{e, \rm max}}_{\gamma_{e, \rm min}}d\gamma_{e}
{dN_{e}\over d\gamma_e}\int^{\infty}_{\nu'_{\rm s,\min}}d\nu'_{\rm
s}\frac{\nu'f'_{\nu'_{\rm s}}}{4\gamma_{e}{\nu'}_{\rm
s}^{2}}g(x,y),
\end{equation}
where $\gamma_{\rm e,\min}=\max[\gamma_{\rm e,c}, \gamma_{\rm e,m},
{h\nu'/( m_{\rm e}c^2)}]$, $\nu'_{\rm s,\min}={\nu'm_{\rm
e}c^2/4[\gamma_{\rm e}(\gamma_{\rm e} m_{\rm e}c^2-h\nu')]}$,
$x=4\gamma_{\rm e} h\nu'_{\rm s}/m_{\rm e}c^{2}$,
$y=h\nu'/[x(\gamma_{\rm e} m_{\rm e}c^{2}-h\nu')]$, and
\begin{equation}
g(x,y)=2y\ln y+(1+2y)(1-y)+\frac{1}{2}{x^{2}y^{2}\over
(1+xy)}(1-y).
\end{equation}

In our case, because the radius of the supernova photosphere is
much smaller than that of the GRB shock, supernova seed photons
can be regarded as a point photon source locating at the center in
the comoving frame of the GRB shock. These soft photons from
supernova impinge on the shock region basically along the radial
direction in the rest frame of the shock, so the scatterings
between these photons and the isotropically-distributed electrons
in the shock are anisotropic. For a photon beam penetrating into
the shock region where the electrons are moving isotropically, the
inverse Compton scattering emissivity of the radiation scattered
at an angle $\theta_{\rm SC}$ relative to the direction of the
photon beam  in the shock comoving frame is (Aharonian $\&$ Atoyan
1981, Brunetti 2000, Fan et. al. 2008):
\begin{equation}\label{AIC}
\begin{array}{ll}\varepsilon'^{\rm AIC}(\nu',{\rm cos}\theta_{\rm SC})=\frac{3\sigma_{\rm
T}c}{16\pi} \int^{\gamma_{e,\rm max}}_{\gamma_{e,\rm
min}}d\gamma_{\rm
e}\frac{dN_{\rm e}}{d\gamma_{e}}\\
\int^{\infty}_{\nu'_{\rm s,min}}\frac{f'^{\rm SN}_{\nu'_{\rm
s}}d\nu'_{\rm s}}{\gamma_{\rm e}^2\nu'_{\rm s}}
[1+\frac{\xi^{2}}{2(1-\xi)}-\frac{2\xi}{b_{\theta}(1-\xi)}+\frac{2\xi^2}{b_{\theta}^2(1-\xi)^2}]\end{array},
\end{equation}
where $\xi\equiv h\nu'/(\gamma_em_{e}c^2)$, $b_{\theta}=2(1-{\rm
cos}\theta_{\rm SC})\gamma_eh\nu'_{\rm s}/(m_ec^2)$ and
$h\nu'_{\rm s}\ll h\nu'\ll \gamma_e
m_ec^2b_{\theta}/(1+b_{\theta})$. On integration over $\theta_{\rm
SC}$ for whole solid angle (i.e. in the case that the photon
distribution is also isotropic), Eq.(\ref{AIC}) reduces to
Eq.(\ref{SSCiso}), i.e. the usual isotropic inverse Compton
scattering emissivity. In the observer frame, the angle $\theta$
between the injecting photons and scattered photons  relates with
the angle $\theta_{\rm SC}$ in the comoving frame by ${\rm
cos}\theta_{\rm SC}=({\rm cos}\theta-\beta)/(1-\beta {\rm
cos}\theta)$, where $\beta$ is the velocity of the GRB shock.

The observed SSC and SNIC flux densities at a frequency $\nu$ are
given respectively by (e.g. Huang et al. 2000, Yu et al. 2007)
\begin{equation}
F_{\nu}^{\rm SSC}=\int^{\pi}_{0}\frac{\varepsilon'^{\rm IC}_{\rm
iso}(\nu/D)}{4\pi D_{\rm L}^2}D^3{\rm sin}\theta d\theta,
\end{equation}
and
\begin{equation}
F_{\nu}^{\rm SNIC}=\int^{\pi}_{0}\frac{\varepsilon'^{\rm
AIC}(\nu/D, {\rm cos}\theta_{\rm SC})}{4\pi D_{\rm L}^2}D^3{\rm
sin}\theta d\theta,
\end{equation}
where $D_{\rm L}$ is the luminosity distance and
$D\equiv[\gamma(1-\beta {\rm cos}\theta)]^{-1}$ is the Doppler
factor.  Note that ${\rm cos}\theta_{\rm SC}$ in Eq.(15) can be
transformed to ${\rm cos}\theta$ through ${\rm cos}\theta_{\rm
SC}=({\rm cos}\theta-\beta)/(1-\beta {\rm cos}\theta)$.

\subsection{Seed Photons from the Supernova}\label{seed photons}
For seed photons from the supernova, we consider the contributions
from two components. One is the early thermal UV-optical emission
from the cooling supernova envelope after being heated by the
radiation-dominated shock (Waxman et al. 2007; Chevalier $\&$
Fransson 2008). Such UV-optical emission  has been observed recently
by {{\it Swift}} UVOT from SN2006aj (Campana et al. 2006) and
SN2008D (Soderberg et al. 2008). Another component is the  supernova
optical emission at later time, powered by the radioactive elements
synthesized in supernovae.

The characterization of the emission from the cooling supernova
envelope is expressed as follows, which is correct at
$t\ga10^2\rm{s}$ (Waxman, M\'{e}sz\'{a}ros, $\&$ Campana 2007):
\begin{equation}\label{rph}
R_{\rm ph}(t)= 3.2\times10^{14}\frac{E_{\rm snej,51}^{0.4}}{(M_{\rm
snej}/M_{\odot})^{0.3}}t_{\rm day}^{0.8}\rm{cm},
\end{equation}
\begin{equation}\label{Tenv}
T_{\rm ph}(t)=2.2\frac{E_{\rm snej,51}^{0.02}}{(M_{\rm
snej}/M_{\odot})^{0.03}}R_{0,12}^{1/4}t_{\rm day}^{-0.5}\rm{eV},
\end{equation}
\begin{equation}\label{Lenv}
L_{\rm ph}(t)=4\pi R_{\rm ph}(t)^{2}\sigma T_{\rm ph}(t)^{4},
\end{equation}
where $R_{\rm ph}(t)$ and $T_{\rm ph}(t)$ are the radius and
temperature of the envelope, $L_{\rm ph}(t)$ is the luminosity from
the photosphere of the cooling envelope, $M_{\rm snej}$ and $E_{\rm
snej}$ are the supernova ejecta mass and energy, and $R_0$ is the
initial stellar radius of the SN progenitor. We take the following
values for supernovae like SN2006aj and SN2008D: $E_{\rm
snej,51}=E_{\rm snej}/(10^{51}\rm{erg})=2$, $M_{\rm
snej}=2M_{\odot}$ and $R_{0,12}=R_{0}/(10^{12}\mbox{cm})=0.3$
(Mazzali et al. 2006; Soderberg et al. 2008).  For hypernovae such
as SN1998bw associated with GRB980425, modelling of the SN optical
emission gives a larger kinetic energy and  ejecta mass such as
$E_{\rm snej,51}\simeq22$ and $M_{\rm snej}=6M_{\odot}$ for SN1998bw
(Woosley et al. 1999).

The luminosity of the late component is rising before $\sim \rm 10
days$ and then  decaying exponentially. For SN2008D, it is found
that the rising is roughly in proportion to $t^{1.6}$ (Soderberg
et. al. 2008), so we can describe the luminosity  as
\begin{equation}\label{Lrad}
L_{\rm rad}(t)=\left\{\begin{array}{ll}3\times10^{42}(\frac{t}{10\rm{days}})^{1.6}\rm{erg~ s^{-1}}, &t<10\rm{days},\\
3\times10^{42}\exp(1-\frac{t}{10\rm{days}})\rm{erg~ s^{-1}},
&t\geq10\rm{days}\end{array}\right .
\end{equation}
We assumed that the radiation temperature is approximately
constant, $T_{\rm rad} \simeq 1\rm{eV}$. The bolometric luminosity
of the cooling supernova envelop is dominant over that of the late
component before $t=7.8\times10^5\rm s$.

Through the calculation of the shock dynamic evolution, the radii
of the GRB shock are $R=1.9\times10^{14}$ $\rm{cm}$ at $t=10^3
\rm{s}$, and $R=1.5\times10^{15}\rm cm$ at $t=10^4\rm s$ for the
trans-relativistic ejecta model, and for the conventional ejecta
model $R=1.0\times 10^{15}\rm cm$ at $t=10^3\rm s$ and
$R=4.1\times 10^{15} \rm cm$ at $t=10^4\rm s$. According to
Eq.\ref{rph}, the shell radii for supernova are $9.6\times
10^{12}\rm cm$ at $t=10^3\rm s$ and $6.1\times 10^{13}\rm cm$ at
$t=10^4\rm s$, and the one for hypernova are $1.8\times 10^{13}\rm
cm$ at $t=10^3\rm s$ and $1.1\times 10^{14}\rm cm$ at $t=10^4\rm
s$. Consequently the shell radius is less than ten percents of the
GRB ejecta height, so it is reasonable to approximate that
supernova photons come from a point source at the center and
impinge onto the electrons in the shock from behind.

\subsection{Compton Parameter $Y$} \label{Y_f}
Following Moderski, Sikora $\&$ Bulik (2000) and Sari $\&$ Esin
(2001), we define the Compton parameter $Y$ as
\begin{equation}\label{Y}
Y\equiv\frac{L_{\rm IC}}{L_{\rm
SYN}}=\frac{{u'}_{\gamma}}{{u'}_{B}}=\frac{{u'}_{\rm SYN}+f_{\rm
a}{u'}_{\rm SN}}{{u'}_{B}},
\end{equation}
where
\begin{eqnarray}
{u'}_{\rm SN}&=&\gamma^{-2}\frac{L_{\rm SN}}{4\pi
cR^2}\nonumber\\
&&\simeq65{\rm~erg~cm^{-3}}t_4^{-0.4}R_{15}^{-2}\gamma^{-2},\\
{u'}_{\rm SYN}&=&{\eta\epsilon_e{u'}\over(1+Y)}\nonumber\\
&&\simeq5.4\times10^2{\rm~erg~cm^{-3}}\epsilon_{e,-0.5}\dot{m}R_{15}^{-2}\gamma(\gamma-1)\eta Y^{-1},\\
{u'}_{B}&=&\epsilon_B{u'}\nonumber\\
&&\simeq1.8{\rm~erg~cm^{-3}}\epsilon_{B,-3}\dot{m}R_{15}^{-2}\gamma(\gamma-1),
\end{eqnarray}
are, respectively, the comoving energy densities of the blackbody
supernova seed photons, synchrotron seed photons and magnetic
fields, and $f_{\rm a}$ is the factor accounting for the
suppression of the photon energy density due to the anisotropic
inverse Compton scattering effect ($f_{\rm a}=1$ corresponds to
the isotropic scattering case). Here $L_{\rm SN}=L_{\rm ph}+L_{\rm
rad}$, which is dominated by the luminosity of the cooling envelop
$L_{\rm ph}$, $u'$ is the comoving internal energy density,
$\eta=\eta_{\rm rad}\eta_{\rm KN}$ is the radiation efficiency
where $\eta_{\rm rad}$ is the fraction that the electron's energy
radiated, and $\eta_{\rm KN}$ is the fraction of synchrotron
photons below the KN limit frequency (Nakar 2007). For slow
cooling, $\eta_{\rm rad}=(\gamma_{e,\rm c}/\gamma_{e,\rm
m})^{2-p}$ (Moderski, Sikora $\&$ Bulik 2000), and
\begin{equation}\label{eta_KN_slow}
\eta_{\rm KN}=\left\{
\begin{array}{ll}
0,&\nu'_{\rm KN}(\gamma_{e,\rm c})\leq \nu'_{\rm m}\\
(\frac{\nu'_{\rm KN}(\gamma_{e,\rm c})}{\nu'_{\rm c}})^{(3-p)/2},&\nu'_{\rm m}< \nu'_{\rm KN}(\gamma_{e,\rm c})<\nu'_{\rm c}\\
1,&\nu'_{\rm c}\leq\nu'_{\rm KN}(\gamma_{e,\rm c})\\
\end{array}\right.
\end{equation}
For fast cooling, $\eta_{\rm rad}=1$ and
\begin{equation}\label{eta_KN_fast}
\eta_{\rm KN}=\left\{
\begin{array}{ll}
0,&\nu'_{\rm KN}(\gamma_{e,\rm m})\leq \nu'_{\rm c}\\
(\frac{\nu'_{\rm KN}(\gamma_{e,\rm m})}{\nu'_{\rm m}})^{1/2},&\nu'_{\rm c}< \nu'_{\rm KN}(\gamma_{e,\rm m})<\nu'_{\rm m}\\
1,&\nu'_{\rm m}\leq\nu'_{\rm KN}(\gamma_{e,\rm m})\\
\end{array}\right.
\end{equation}

 Solving eq. (\ref{Y}), we get
\begin{equation}\label{Y_eta}
Y={1\over2}\left[\sqrt{4\frac{\eta\epsilon_{e}}{\epsilon_{B}}+\left(1+\frac{f_{\rm
a}{u'}_{\rm SN}}{\epsilon_{B}{u'}}\right)^{2}} +\left(\frac{f_{\rm
a}{u'}_{\rm SN}}{\epsilon_{B}{u'}}-1\right)\right].
\end{equation}
Roughly, the above expression can be simplified in three limiting
cases as follows
\begin{equation}\label{Y_eta}
Y=\left\{
\begin{array}{ll}
f_{\rm a}u'_{\rm SN}/(\epsilon_Bu'),&f_{\rm
a}{u'}_{\rm SN}\gg u'_{\rm SYN}\\
\sqrt{{\eta\epsilon_{e}/\epsilon_{B}}},&u'_{B}\ll f_{\rm a}{u'}_{\rm
SN}\ll u'_{\rm SYN} \\
(\sqrt{4{\eta\epsilon_{e}/\epsilon_{B}}+1}-1)/2,&f_{\rm a}{u'}_{\rm
SN}\ll {\rm min}[{u'}_{\rm SYN},u'_{B}]\\
\end{array}\right.
\end{equation}
In the latter two cases, $Y$ can be treated as a constant when the
electrons are in the fast-cooling regime and $\nu'_{\rm
m}\leq\nu'_{\rm KN}(\gamma_{e,\rm m})$ with $\eta=1$.

\subsection{Pair Production Opacity For High Energy Photons}\label{tao}
High-energy gamma-rays can be attenuated due to interaction with
low-energy photons through the pair production effect. We consider
the pair production opacity in the shock frame due to the absorption
by low-energy photons, which include thermal photons from the
supernova, synchrotron photons, SSC photons and SNIC photons. A high
energy photon of energy $E'_{\gamma,1}$ in the shock frame will
annihilate with a low energy photon of $E'_{\gamma,2}$, provided
that
$E'_{\gamma,1}E'_{\gamma,2}(1-\cos\theta_{12})\geq2(m_ec^{2})^{2}$,
where $\theta_{12}$ is the collision angle of the two annihilation
photons. The pair creation cross section is given by
\begin{equation}
\sigma(E'_{\gamma,1},E'_{\gamma,2})=\frac{1}{2}\pi
r_{0}^{2}(1-\tilde{\beta}^{2})[(3-\tilde{\beta}^{4}){\rm
ln}\frac{1+\tilde{\beta}}{1-\tilde{\beta}}-2\tilde{\beta}(2-\tilde{\beta}^2)],
\end{equation}
where $\tilde{\beta}\equiv
v/c=\sqrt{1-2(m_{e}c^{2})^{2}/[E'_{\gamma,1}E'_{\gamma,2}(1-\cos\theta_{12})}]$
is the velocity of electrons in the center-of-mass frame (Heitler
1954, Stecker, De Jager, $\&$ Salamon, 1992) and $r_0=e^2/(m_ec^2)$
is the classic electron radius. By using the photon distribution in
the shock frame, we obtain the optical depth for high-energy photons
of energy,
\begin{equation}
\tau(E'_{\gamma,1})=\int_{E'_{thr}}^{\infty}\sigma(E'_{\gamma,1},E'_{\gamma,2})n_{\gamma}(E'_{\gamma,2})\frac{R}{\eta_{\rm
R}}dE'_{\gamma,2},
\end{equation}
where $\eta_{\rm R}$ is the shock compressed ratio, which is
$\eta_{\rm R}=4\gamma+3$, and the threshold energy of low energy
photons is $E'_{\rm thr}=2(m_{e}c^{2})^{2}/E'_{\gamma,1}(1-
\cos\theta_{12})$. In the calculation, we estimate the cutoff energy
conservatively by assuming that colliding photons are moving
isotropically in the shock frame. Such a treatment may overestimate
the pair-production opacity because in reality the supernova seed
photons move anisotropically (i.e. moving outward in radial
direction as seen by high-energy photons emitted from the shock at
much larger radii).

\begin{figure}
\epsscale{.80} \plotone{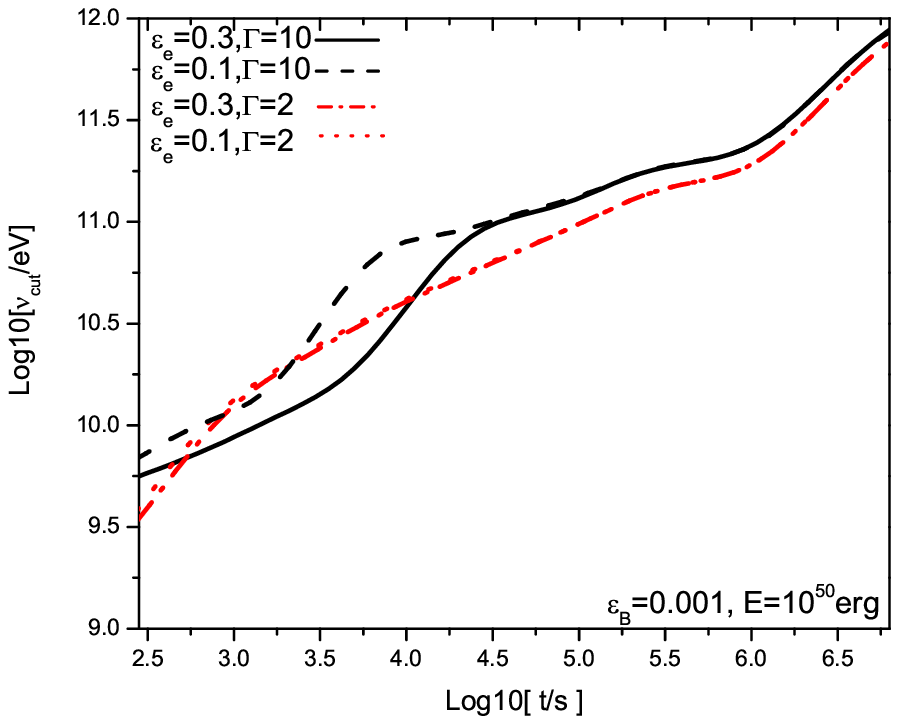} \caption{The high-energy cutoff
due to annihilation with low-energy photons at different times in
the two models for low-luminosity GRBs, with $\epsilon_B=0.001$,
$p=2.2$, $\dot{m}=1$, $E=10^{50}\rm{erg}$ and different parameters
of $\epsilon_e$. The solid lines and dashed lines show the cutoff
energy in the conventional relativistic ejecta model with
$\epsilon_e=0.3$ and $\epsilon_e=0.1$, respectively. The
dash-dotted and dotted lines show the cutoff energy in the
trans-relativistic ejecta model with $\epsilon_e=0.3$ and
$\epsilon_e=0.1$, respectively.\label{nucut}}
\end{figure}

The result of the cutoff energy in the observer frame
$E_{\gamma,{\rm cut}}=\gamma E'_{\gamma}(\tau=1)$ is given in Figure
\ref{nucut}. Since the luminosity and peak energy of the supernova
envelope emission decreases with time in general and the shock
radius increases with time, the cutoff energy of the high energy
spectrum increases with time, which is clearly seen in Fig.
\ref{nucut}. From this figure, we can see that the cutoff energy is
above $1\rm{GeV}$ after the starting time of our calculation
($10^{2.5}\rm{s}$) , so we can calculate the light curves at energy
$\sim 1\rm{GeV}$ without considering the opacity. Since the cutoff
energy is large enough, it does not affect the detectability of
Fermi LAT, whose sensitive energy band is
$20\rm{MeV}\sim300\rm{GeV}$.

\section{Results}
\subsection{Spectra and Light Curves: Numerical Results}\label{spectra and light curves}
\begin{figure}
\epsscale{.80} \plotone{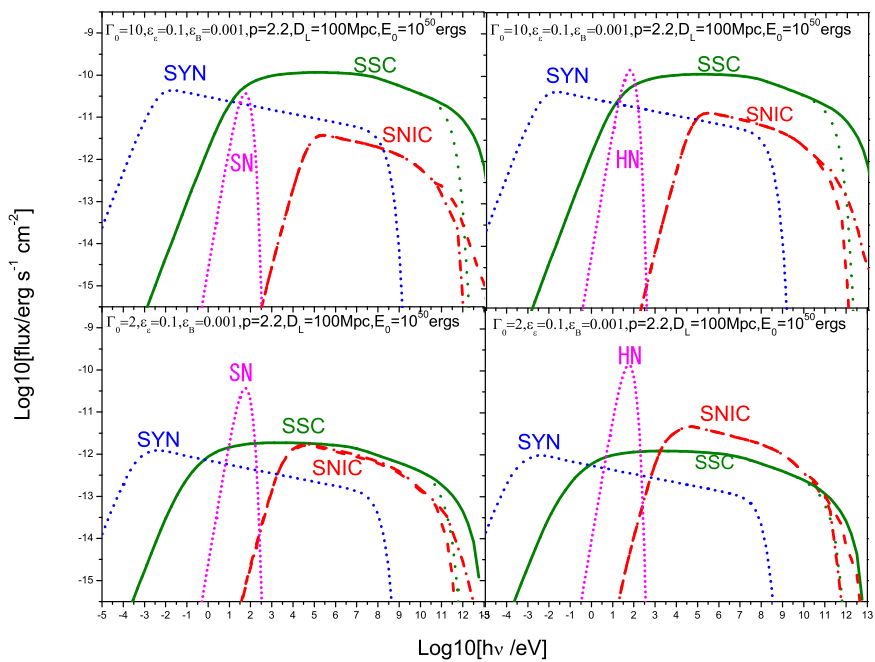} \caption{The spectra of SNIC, SSC,
synchrotron emission and supernova/hypernova photons in the two
models at times $t=10^3\rm{s}$ for the parameters
$\epsilon_e=0.1$, $\epsilon_e=0.001$, $p=2.2$ and
$E=10^{50}\rm{erg}$. The left panels show the supernova case, the
right ones show the hypernova case, the top ones denote the
conventional relativistic ejecta model and the bottom ones denote
the trans-relativistic ejecta model. The dotted and dash dotted
lines show the spectra of SSC and SNIC with the annihilation
effect taken into account, while the solid lines and dashed lines
show the spcetra of SSC and SNIC emission without considering this
effect. The short dashed lines denote the spectra of synchrotron
emission and the short dotted lines denote the blackbody spectrum
from the supernova/hypernova. \label{spectra_13}}
\end{figure}

We first compare the spectra of different IC components for the two
models. The spectra at $t=10^3\rm{s}$ are shown in Figure
\ref{spectra_13}, including the synchrotron emission, the thermal
emission from the supernova or hypernova, the synchrotron
self-Compton emission and the SNIC emission, for parameters
$\epsilon_e=0.1$, $\epsilon_B=0.001$, $p=2.2$, $E=10^{50}\rm{erg}$
and burst distance $D_L=100\rm{Mpc}$. From the spectra we can see
that for the supernova case the SSC emission dominates over the SNIC
emission at energies from $\rm 1 MeV$ to $\rm 100 GeV $ for
conventional relativistic ejecta model, and for trans-relativistic
ejecta model the SSC emission is higher than SNIC emission and the
two components are both important at high energies below the cutoff
energy. For the hypernova case the SSC emission is dominant in the
conventional relativistic ejecta model while the SNIC emission is
dominant in the trans-relativistic ejecta model at the high energy
band. This is because in the case of hypernovae and $\Gamma_0=2$,
the energy density of hupernova photons $u'_{\rm HN}$ (multiplied by
the suppressed factor $f_{\rm a}$ due to the anisotropic scattering)
is significantly higher than the synchrotron radiation density
$u'_{\rm SYN}$.

\begin{figure}
\epsscale{.80} \plotone{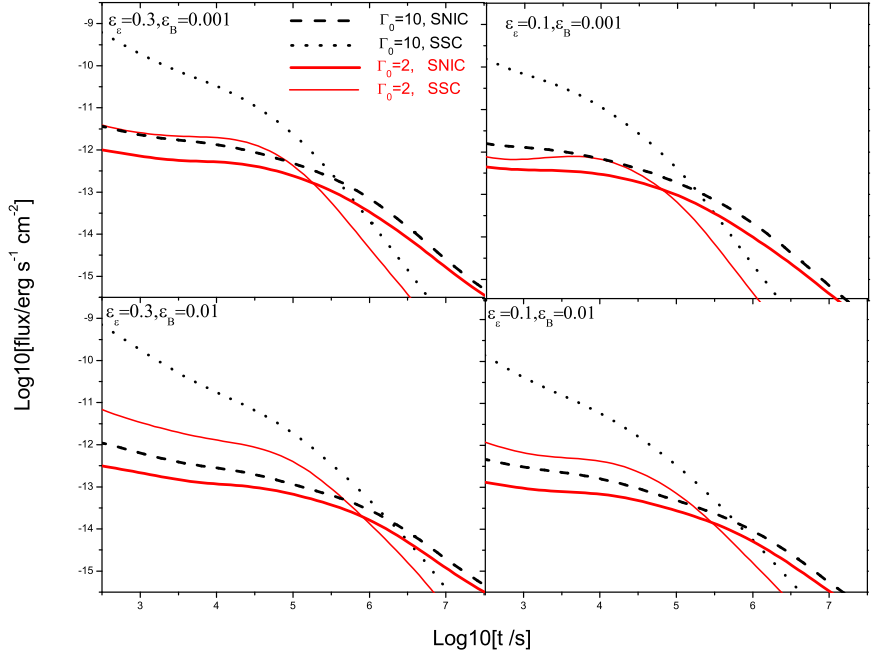} \caption{Light curves of the SNIC
and SSC emission at energy $h\nu=\rm 1GeV$ in the two models for a
normal supernova case and for $E=10^{50}\rm{erg}$ and different
values for $\epsilon_e$ and $\epsilon_B$. The thicker and thinner
solid lines represent the SNIC and SSC emission in the
trans-relativistic ejecta model, and the dashed and the dotted
lines represent the SNIC and SSC emission in the conventional
relativistic ejecta model. \label{lc_sup}}
\end{figure}

Figure \ref{lc_sup} shows light curves of SSC and SNIC afterglow
emission at $h\nu=\rm{1 GeV}$ in the two models for
$E=10^{50}\rm{ergs}$, $p=2.2$ and four different sets of parameters
of $\epsilon_e$ and $\epsilon_B$ for the supernova case. In the
conventional relativistic model with $\Gamma_0=10$, a sharp decay
phase of a GeV afterglow is produced during the early hours, which
is mildly dominated by the SSC emission, and a slightly flatter
decay phase dominated by SNIC emission takes over at late times
$t>10^6\rm s$. On the other hand, for the trans-relativistic ejecta
model with $\Gamma_0=2$,  a plateau of SSC emission, due to the
presence of a coasting phase in the ejecta dynamic, dominates in the
early time, which transits to a faster decay at later time and SNIC
emission become dominant after the time $10^5\sim10^6\rm s$.

\begin{figure}
\epsscale{.80} \plotone{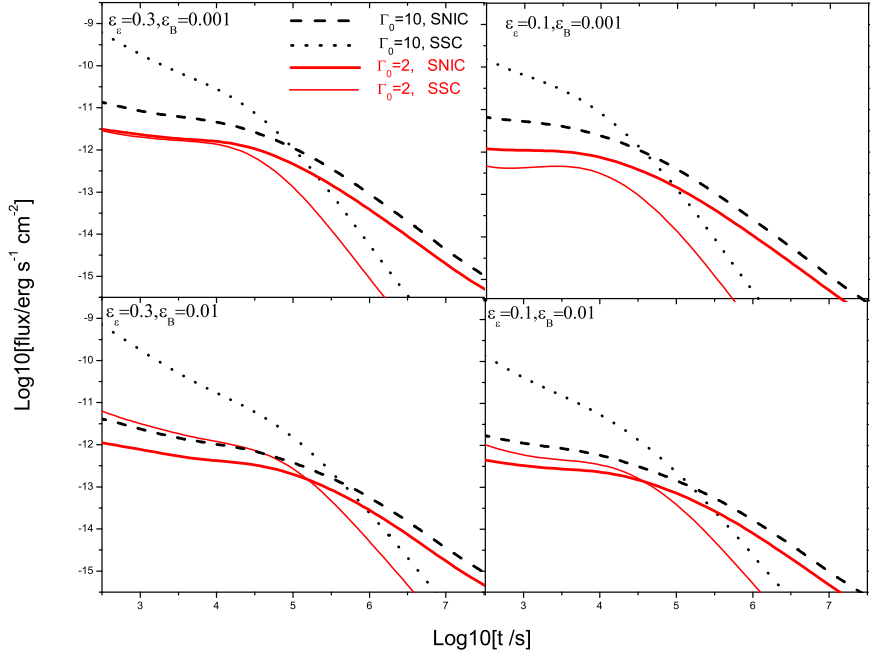} \caption{Same as Fig.3, but for
the hypernova case. The thicker and thinner solid lines represent
the SNIC and SSC emission in the trans-relativistic ejecta model,
and the dashed and the dotted lines represent the SNIC and SSC
emission in the conventional relativistic ejecta model.
\label{lc_hyp}}
\end{figure}

Figure \ref{lc_hyp} show light curves for the hypernovae case. Light
curves in the conventional relativistic ejecta model are similar
with those for the supernova case except that SNIC emission become
dominant at earlier time; for the trans-relativistic ejecta model,
SNIC emission is always dominated with parameters $\epsilon_e=0.1$
and $\epsilon_B=0.001$, a plateau is seen in the early time, which
transits to a faster decay at later time.

 The two models also predict different flux levels at
high-energies. In early hours, the total flux from the conventional
relativistic ejecta model is more than one order of magnitude higher
than that in the trans-relativistic ejecta model. This can be
explained by the different amount of energy in shocked electrons.
For the conventional relativistic ejecta model, the ejecta has been
decelerated at this time and a great part of its energy has been
converted into shocked electrons, while in the trans-relativistic
ejecta model, only a small fraction of the ejecta kinetic energy has
been converted to shocked electrons at this early time. A lower
amount of energy in shocked electrons results in a lower flux level
in the trans-relativistic ejecta model.

In addition, a comparison among four panels in Figure \ref{lc_sup}
and Figure \ref{lc_hyp} indicates that the flux decreases as
$\epsilon_e$ decreases, which is obvious since the energy of
radiating electrons $E_e\propto \epsilon_e$. As $\epsilon_B$
decreases, the SSC flux changes little and the SNIC flux increases
for the case that SSC emission is dominant at early time. This can
be understood from the following analysis. Since the SSC emission
dominates and $f_{\rm a}u'_{\rm SN}\gg u'_{B}$ at the early times,
$Y\simeq \sqrt{\eta
\epsilon_e/\epsilon_B}\propto\epsilon_B^{-\frac{1}{2}}\epsilon_e^{\frac{1}{2}}$.
The SSC and SNIC flux at $\nu>\nu_{\rm min}>\nu_{\rm c}$ scale as
\begin{equation}
\nu F_{\nu}^{\rm SSC}=\nu F_{\rm max}^{\rm SSC}(\frac{\nu_{\rm
min}^{\rm SSC}}{\nu_{c}^{\rm
SSC}})^{-\frac{1}{2}}(\frac{\nu}{\nu_{\rm min}^{\rm
SSC}})^{-\frac{p}{2}}\propto\epsilon_B^{\frac{p}{4}-\frac{1}{2}}\epsilon_e^{2p-3}
\end{equation}
and
\begin{equation}
\nu F_{\nu}^{\rm SNIC}=\nu F_{\rm max}^{\rm SNIC}(\frac{\nu_{\rm
min}^{\rm SNIC}}{\nu_{c}^{\rm
SNIC}})^{-\frac{1}{2}}(\frac{\nu}{\nu_{\rm min}^{\rm
SNIC}})^{-\frac{p}{2}}\propto\epsilon_B^{-\frac{1}{2}}\epsilon_e^{p-\frac{3}{2}}.
\end{equation}

\subsection{The effect of the anisotropic scattering on the SNIC emission}\label{aiciso}
\begin{figure}
\epsscale{.80} \plotone{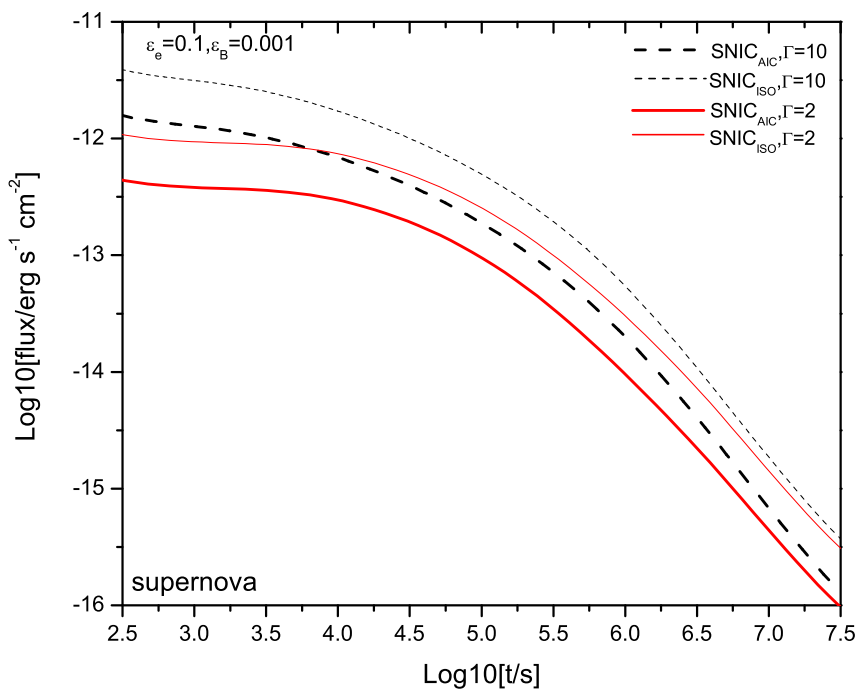} \caption{Comparison of the light
curves of the SNIC emission at $h\nu=\rm 1GeV$  with (the thicker
lines) and without (the thinner lines) the anisotropic scattering
effect correction in the two models of low-luminosity GRBs. The
solid lines represent the SNIC emission in the trans-relativistic
ejecta model, and the dashed lines represent the SNIC emission in
the conventional relativistic ejecta model. \label{lc_iso}}
\end{figure}

The incoming supernova photons are  anisotropic as seen by the
isotropically distributed electrons in the GRB shock, so the IC
scatterings are anisotropic. In order to see how  the anisotropic
inverse-Compton scattering (AIC)  affects the SNIC flux, we
compare the light curves of the SNIC emission obtained by using
the isotropic scattering formula Eq.11 and using the AIC
scattering formula Eq.13 in Figure \ref{lc_iso}. The thinner lines
show the SNIC light curves obtained using the usual isotropic
scattering formula, while the thicker ones correspond to the
calculations with the AIC scattering effect taken into account.
One can see that the flux of the SNIC emission with the AIC effect
correction is reduced by a factor of about $\sim0.4$ compared to
the isotropic scattering case. This is consistent with the
calculation result obtained by Fan $\&$ Piran (2006), who studied
the anisotropic inverse Compton scattering between inner
optical/X-ray flare photons and electrons in the outer GRB forward
shock.

Fig.5 shows that the AIC effect suppresses the SNIC flux only
slightly. The anisotropic photon distribution results in more
head-on scatterings, i.e. the photon beam scatter preferentially
with those electrons that move in the direction antiparallel to
the photon beam, so one can expect that the scattered IC emission
power has a maximum at $\theta_{\rm SC}=\pi$ and goes to zero for
small scattering angles (e.g. Brunetti 2000). The photons
scattered into the angles $0\la\theta_{\rm SC}\la\pi/2$ relative
to the shock moving direction in the shock comoving frame will
fall into the cone of angle $1/\Gamma$ in the observer frame,
according to the transformation formula ${\rm cos}\theta_{\rm
SC}=({\rm cos}\theta-\beta)/(1-\beta {\rm cos}\theta)$. Therefore
the AIC scatterings decrease the IC emission in the $1/\Gamma$
cone along the direction of the photon beam, but meanwhile they
enhance the emission at larger angles (about half of the emission
falling into angles between $1/\Gamma$ and $2/\Gamma$, see Wang \&
\Meszaros 2006). For a spherical outflow as we consider here, the
IC emission after integration over angles should have the same
flux in every direction in the observer frame, with a flux level
only slightly reduced comparable to the isotropic scattering case.

\subsection{Analytical Light Curves}\label{light curves}

As a comparison, we derive here approximate analytical expressions
for afterglow light curves, which provide an explanation for the
physical origin of the behavior. Since the anisotropic SNIC emission
are depressed by a factor of $~0.4$, which is almost constant,
relative to the isotropic seed photons case, we can consider the
isotropic seed photons case for the approximate analytic treatment
of the afterglow light curves. The blackbody photons from the
supernova can be approximated as mono-energetic photons with
$h\nu_{\rm SN}=2.7 \rm{K}T_{\rm SN}$. Thus, similar to the
description for synchrotron emission of a single electron (Sari et
al. 1998), the radiation power and characteristic frequencies of
SNIC from a single electron scattering supernova photons in the
observer frame can be described by
\begin{equation}
P(\gamma_{e})=\frac{4}{3}\sigma_Tc\gamma^2\gamma_e^2\frac{L_{\rm{SN}}}{\gamma^2\pi
R^2c},
\end{equation}
and
\begin{equation}
\nu(\gamma_{e})=2\gamma\gamma_e^2\nu_{\rm{SN}}/\gamma=2\gamma_e^2\nu_{\rm{SN}},
\end{equation}
respectively. Similar to the analysis in Sari and Esin (2001), the
maximum flux of the SNIC spectrum is
\begin{equation}
F_{\rm max}^{\rm SNIC}=\frac{N_e}{4\pi
D_{\rm L}^2}\frac{P(\gamma_{e})}{\nu(\gamma_{e})} 
\end{equation}
Characteristic SNIC frequencies are
\begin{equation}\label{numinsnic}
\nu_{\rm min}^{\rm SNIC}=2\gamma_{e,\rm m}^2\nu_{\rm SN}
\end{equation}
and
\begin{equation}
\nu_{c}^{\rm SNIC}=2\gamma_{e,\rm c}^2\nu_{\rm SN}
\end{equation}
respectively.

By adopting the broken power-law approximation for the IC spectral
component ( Sari \& Esin 2001) and the dynamics of the shock
discussed in $\S$ \ref{models}, one can derive the analytic light
curves in an approximate way.

\subsubsection{Light Curves In The Conventional Relativistic Ejecta
Model} In the conventional relativistic ejecta model, at time
$t_{\rm dec}<t<t_{\rm tran}$, where $t_{\rm tran}$ is defined as the
transition time when $\gamma_{e,\rm m}=\gamma_{e,\rm c}$, we have
$\eta=1$ and $u'_{B}\ll f_{\rm a}u'_{\rm SN}\ll u'_{\rm SYN}$, so
$Y\simeq\sqrt{\eta\epsilon_e/\epsilon_B}\propto t^{0}$.
 The shock
dynamic follows  the Blandford \& McKee self-similar solution in the
wind medium, i.e. $R\propto t^{\frac{1}{2}}$ and $\gamma\propto
t^{-\frac{1}{4}}$. Then we can obtain  the evolution of the break
frequencies of the SSC and SNIC spectral components and their peak
flux  in the following way:
\begin{equation}
\nu_{\rm min}^{\rm SSC}=2\gamma_{e,\rm m}^2\nu_{\rm min}\propto
t^{-2},
\nu_{\rm c}^{\rm SSC}=2\gamma_{e,\rm c}^2\nu_{\rm c}\propto t^{2},\\
\end{equation}
\begin{equation}
F_{\nu,\rm max}^{\rm SSC}=\frac{\sigma_TN_e}{4\pi R^2}F_{\rm
max}^{\rm SYN}\propto t^{-1}.
\end{equation}
and
\begin{equation}
\nu_{\rm min}^{\rm SNIC}\propto t^{-1}, \nu_{\rm c}^{\rm
SNIC}\propto t, F_{\nu,\rm max}^{\rm SNIC}\propto t^{-0.4}.
\end{equation}
where $\nu_{\rm SN}\propto T_{\rm SN}\propto t^{-1/2}$ has been used
for the cooling envelope emission (see Eq. 13). The SSC and SNIC
flux at an observed frequency $\nu$ higher than characteristic
frequencies vary as $\nu F_{\nu}^{\rm SSC}\propto t^{-p+1}$ and $\nu
F_{\nu}^{\rm SNIC}\propto t^{-\frac{p}{2}+0.6}$ respectively for
$t<t_{\rm tran}$.

At $t>t_{\rm SNIC}$, where $t_{\rm SNIC}$ is the time when $f_{\rm
a}u'_{\rm SN}=u'_{\rm SYN}$, we take $Y\simeq f_{\rm a}{u'}_{\rm
SN}/(\epsilon_{B}{u'})$ because the SNIC emission becomes dominated.
At such time, the shock is likely to enter the non-relativistic
phase, so we take the Sedov-von Neumann-Taylor solutions $R\propto
t^{\frac{2}{3}}$ and $\beta\propto t^{-\frac{1}{3}}$, which induce
that $Y\propto t^{\frac{2}{3}-0.4}$. So the minimum and cooling
Lorentz factors of electrons vary as $\gamma_{e,\rm m}\propto
\beta^{2}\propto t^{-\frac{2}{3}}$ and $\gamma_{e,\rm
c}\propto\beta^{-2}R^2t^{-1}Y^{-1}\propto t^{0.4+\frac{1}{3}}$.
Thus, the break frequencies of the SSC and SNIC spectral components
and their peak fluxes evolve with time in the following way:
\begin{equation}
\nu_{\rm min}^{\rm SSC}\propto t^{-\frac{11}{3}}, \nu_{\rm c}^{\rm
SSC}\propto t^{\frac{1}{3}+1.6}, F_{\nu,\rm max}^{\rm SSC}\propto
t^{-1},
\end{equation}
and
\begin{equation}
\nu_{\rm min}^{\rm SNIC}\propto t^{-\frac{11}{6}}, \nu_{\rm c}^{\rm
SNIC}\propto t^{\frac{1}{6}+0.8}, F_{\nu,\rm max}^{\rm SNIC}\propto
t^{-\frac{2}{3}+0.1}.
\end{equation}
Then the SSC and SNIC fluxes at high energy $\nu$ vary as $\nu
F_{\nu}^{\rm SSC}\propto t^{1.8-\frac{11}{6}p}$ and $\nu
F_{\nu}^{\rm SNIC}\propto t^{\frac{5}{6}-\frac{11}{12}p}$ at
$t>t_{\rm SNIC}$.

To summarize, the temporal evolution of the SSC and SNIC afterglow
emission at high energies are
\begin{equation} \nu
F_{\nu}^{\rm SSC}\propto\left \{\begin{array}{ll}
t^{-1.2}&t\la t_{\rm tran}\\
t^{-2.2}&t>t_{\rm SNIC}\end{array}\right.
\end{equation}
and
\begin{equation}
\nu F_{\nu}^{\rm SNIC}\propto\left \{\begin{array}{ll}
t^{-0.5}&t\la t_{\rm tran}\\
t^{-1.2}&t>t_{\rm SNIC}\end{array}\right.
\end{equation}
in the two asymptotic phases for $p=2.2$.

\subsubsection{Light Curves In The Trans-relativistic Ejecta Model}
In the trans-relativistic ejecta model, one would expect a much
flatter light curve of high-energy gamma-ray emission at early
times, which could be dominated by both the SSC emission and SNIC
emission, depending on the properties of the underlying supernova
and the shock parameter $\epsilon_e$ and $\epsilon_B$. For the
supernova case, the SSC emission is dominant before the deceleration
time and the transition time for most parameters, while for the
hypernova case, the SSC emission is dominant at earlier time with
$\epsilon_e=0.3$ and $\epsilon_B=0.01$ and the SNIC emission is
always dominant for $\epsilon_e=0.1$ and $\epsilon_B=0.001$.

For cases where SSC emission dominated in trans-relativistic ejecta
model, we have $\eta=1$ in fast cooling regime and $u'_{B}\ll f_{\rm
a}u'_{\rm SN}\ll u'_{\rm SYN}$, so
$Y\simeq\sqrt{\eta\epsilon_e/\epsilon_B}\propto t^{0}$. Since
$t<t_{\rm dec}$, we adopt the approximation $R\propto t$,
$\gamma\propto t^0$. So the SSC and SNIC flux vary as $\nu
F_{\nu}^{\rm SSC}\propto t^{1-\frac{p}{2}}$ and $\nu F_{\nu}^{\rm
SNIC}\propto t^{0.1-\frac{p}{4}}$ respectively at $t<{\rm
min}(t_{\rm dec},t_{\rm tran})$. At later time when $t>t_{\rm dec}$
and $\gamma\rightarrow1$, the SNIC emission become dominant, the
evolution of light curves is the same as that in conventional
relativistic ejecta model, i.e. the SSC and SNIC flux vary as $\nu
F_{\nu}^{\rm SSC}\propto t^{1.8-\frac{11}{6}p}$ and $\nu
F_{\nu}^{\rm SNIC}\propto t^{\frac{5}{6}-\frac{11}{12}p}$ at
$t>t_{\rm dec}$. Therefore, for the supernova case and the hypernova
case with early dominated SSC emission, the temporal evolution of
the SSC and SNIC emission at high frequency $\nu$ for $p=2.2$ are
given by
\begin{equation}
\nu F_{\nu}^{\rm SSC}\propto\left \{\begin{array}{ll}
t^{-0.1}&t\la{\rm min}(t_{\rm dec}, t_{\rm tran})\\
t^{-2.2}&t\gg t_{\rm dec}\end{array}\right.
\end{equation}
and
\begin{equation}
\nu F_{\nu}^{\rm SNIC}\propto\left \{\begin{array}{ll}
t^{-0.45}&t\la{\rm min}(t_{\rm dec}, t_{\rm tran})\\
t^{-1.2}&t\gg t_{\rm dec}\end{array}\right.
\end{equation}.

For cases where SNIC emission dominated in trans-relativistic ejecta
model, We have $f_{\rm a}u'_{\rm SN}\gg u'_{\rm SYN}$, so $Y\simeq
f_{\rm a}u'_{\rm SN}/(\epsilon_Bu')$. At the time $t<t_{\rm dec}$,
we adopt the approximation $R\propto t$, $\gamma\propto t^0$ to
yield $Y\propto t^{-0.4}$. So the SSC and SNIC flux vary as $\nu
F_{\nu}^{\rm SSC}\propto t^{1.8-\frac{p}{2}}$ and $\nu F_{\nu}^{\rm
SNIC}\propto t^{0.5-\frac{p}{4}}$ respectively at $t<{\rm
min}(t_{\rm dec},t_{\rm tran})$. At later time when $t>t_{\rm dec}$
and $\gamma\rightarrow1$, the evolution of light curves is the same
as that in conventional relativistic ejecta model, i.e. the SSC and
SNIC flux vary as $\nu F_{\nu}^{\rm SSC}\propto
t^{1.8-\frac{11}{6}p}$ and $\nu F_{\nu}^{\rm SNIC}\propto
t^{\frac{5}{6}-\frac{11}{12}p}$ at $t>t_{\rm dec}$. Therefore, for
the supernova case and the hypernova case with early dominated SSC
emission, the temporal evolution of the SSC and SNIC emission at
high frequency $\nu$ for $p=2.2$ are given by
\begin{equation}
\nu F_{\nu}^{\rm SSC}\propto\left \{\begin{array}{ll}
t^{0.7}&t\la{\rm min}(t_{\rm dec}, t_{\rm tran})\\
t^{-2.2}&t\gg t_{\rm dec}\end{array}\right.
\end{equation}
and
\begin{equation}
\nu F_{\nu}^{\rm SNIC}\propto\left \{\begin{array}{ll}
t^{-0.05}&t\la{\rm min}(t_{\rm dec}, t_{\rm tran})\\
t^{-1.2}&t\gg t_{\rm dec}\end{array}\right.
\end{equation}.

\subsection{Detectability by the Fermi LAT}\label{detectability of Fermi LAT}
We explore here whether Fermi LAT can detect the  high energy
gamma-ray emission from low luminosity GRBs in the two models
considered above. Following Zhang $\&$ M\'{e}sz\'{a}ros (2001),
Gou $\&$ M\'{e}sz\'{a}ros (2007) and Yu, Liu $\&$ Dai (2007), the
fluence threshold for long-duration observations is $F_{\rm
thr}=[\phi_0(t/t_{\rm eff})^{1/2}]E_{\rm ph}t_{\rm eff}$ which is
in proportional to $t^{1/2}$ due to the limitation by the
background, where we take the average energy of the detected
photons as $E_{\rm ph}=400\rm MeV$ and the effective time as
$t_{\rm eff}=0.5\rm yr$. $\phi_0$ is the integral sensitivity
above 100 $\rm MeV$ for LAT for a steady source after a year sky
survey, which is $\phi_0\sim3\times10^{-9}\rm phs cm^{-2}s^{-1}$
(atwood et. al. 2009) and is improved by a factor of $3$ by
keeping the GRB position at the center of the LAT field of view as
long as possible (Gou $\&$ M\'{e}sz\'{a}ros 2007). For short-time
observation, the fluence threshold is calculated by $F_{\rm
thr}=5E_{\rm ph}/A_{\rm eff}$ under the assumption that at least
$5$ photons are collected. Taking the effective area $A_{\rm
eff}=6000\rm cm^2$, we can obtain the fluence threshold of Fermi
LAT,
\begin{equation}
F_{\rm thr}=\left\{\begin{array}{ll}
5.3\times10^{-7} \rm{erg~cm^{-2}},&t\leq4.4\times10^{4}\rm{s},\\
2.5\times10^{-9}t^{1/2} \rm{erg~cm^{-2}},&t>4.4\times10^{4}\rm{s}
.\end{array}\right .
\end{equation}
With this fluence threshold, the detectability of high energy
emission (with supernova seed photons luminosity given above and a
total energy $E=10^{50}\rm{erg}$) by the Fermi LAT is shown in
Figure 6. The time-integrated fluence shown in the plot is defined
as an integration of the flux density ($F_{\nu}$) over the Fermi
LAT energy band $[20\rm MeV, 300\rm GeV]$ and the time interval
$[0.5t,t]$ as used in Gou $\&$ Meszaros (2007), which is
$\int_{0.5t}^{t}\int_{\nu_{1}}^{\nu_{2}}F_{\nu}d\nu dt$.

For $\epsilon_e=0.3$ and other representative parameter values,
high-energy gamma-ray emission in the conventional relativistic
ejecta model stays detectable up to $\sim10^6\rm{s}$, while the
high-energy gamma-ray emission in the trans-relativistic ejecta
model can only be detected in a short period around
$10^{4.5}\rm{s}$. For a lower value such as $\epsilon_e=0.1$, the
high-energy gamma-ray emission can still be detected in the
conventional relativistic ejecta model, while it becomes
undetectable for the trans-relativistic ejecta model.

In order to compare with earlier results in Ando \&
M\'{e}sz\'{a}ros (2008), we increase the total energy to
$E=2\times10^{50}\rm{erg}$, which yields  a  kinetic energy
$E_{\rm k}=(\Gamma_0-1)/\Gamma_0 E=10^{50}\rm{erg}$ in the
trans-relativistic ejecta model, the same as that used in Ando \&
M\'{e}sz\'{a}ros (2008). This will increase the  fluence by a
factor of $2$. We also increase the SN luminosity from that of a
normal SN  (shown in $\S$ \ref{seed photons}) to SN1998bw-like
hypernovae. In Figure 7, we show the detectability of high energy
emission by Fermi LAT in this case. By comparing the flux of light
curves between the supernova case and the hypernova case, which is
shown in Fig. \ref{lc_sup} and Fig. \ref{lc_hyp}, we can see that
increasing the SN luminosity can hardly enhance the total IC flux
for the two models. Our IC flux is still lower than that obtained
by Ando \& \Meszaros (2008). The main difference between our
calculation and that of Ando \& M\'{e}sz\'{a}ros (2008) is the
different minimum Lorentz factors used in the calculations. Ando
\& M\'{e}sz\'{a}ros (2008) may overestimate the minimum Lorentz
factor of electrons by taking $\gamma_{e,\rm
m}=\epsilon_e(m_p/m_e)\gamma$, which is a factor of $(p-1)/(p-2)$
larger than ours. From the formula describing the high energy flux
$\nu F_{\nu}^{\rm SNIC}\propto\nu_{\rm min,\rm
SNIC}^{(p-1)/2}\propto\gamma_{e,\rm m}^{p-1}$, one expects that
the flux is increased by a factor of $(\frac{p-1}{p-2})^{p-1}$,
which is about $8$ for $p=2.2$.

\section{Discussions and Conclusions}
The external stellar wind provides a source of Thomson opacity to
scatter the supernova emission, thus a quasi-isotropic,
back-scattered SN radiation field is present.  Let's study whether
this component is important. For a GRB shock locating at radius
$R$ and moving with a Lorentz factor $\gamma$, the Thomson
scattering optical depth of the wind is $\tau_{\rm
w}=\sigma_TRn=\sigma_TKR^{-1}=2.0\times10^{-4}\dot{m}R_{15}^{-1}$.
The scattered SN energy density by the wind in the comoving frame
of the blast wave is $u'^{\rm w}_{\rm SN}=(L_{SN}/{4\pi c R^2})
\tau_{\rm w} \gamma^2$ due to the relativistic boosting effect. On
the other hand, the energy density of the supernova photons
impinging the shock from behind is $u'_{\rm
SN}=\gamma^{-2}(L_{SN}/{4\pi c R^2})$ (see Eq.21). The ratio
between these two energy densities is $u'^{\rm w}_{\rm SN}/u'_{\rm
SN}=2.0\times10^{-4}\dot{m}R_{15}^{-1} \gamma^4$. For the
trans-relativistic ejecta model with $\Gamma_0=2$, the radius of
the GRB ejecta is $R=1.9\times10^{14}$ $\rm{cm}$ at $t=10^3
\rm{s}$, so $u'^{\rm w}_{\rm SN}\ll u'_{\rm SN}$ for typical wind
parameters. For the conventional relativistic ejecta model, the
Lorentz factor and radius of the GRB ejecta are respectively
$\gamma\sim4$ and $R=1.0\times10^{15}$ $\rm{cm}$ at $t=10^3
\rm{s}$, so we also have $u'^{\rm w}_{\rm SN}\ll u'_{\rm SN}$.
This means that the wind-scattered supernova radiation field is
subdominant compared to the direct impinging supernova photon
field and hence we neglect its contribution to the high-energy
gamma-ray emission.

Our estimate of the pair-production opacity for high-energy
photons in $\S$ \ref{tao} is based on the common assumption that
the colliding photons are isotropic in the rest frame. However, as
we have shown above, the low-energy photons from the supernova
essentially move radially outward before colliding with
high-energy photons (e.g. Wang, Li \& \Meszaros 2006). Therefore,
the collision process between the high-energy photons and soft
supernova photons is anisotropic, which would decrease the
pair-production opacity. This will be the subject of a more
detailed future calculation and would be useful to explore whether
$\rm{TeV}$ photons can escape from the source, which is important
for checking the detectability by ground-based Cherenkov detectors
such as Magic, VERITAS, Milago, HESS, ARGO etc. Additionally,
besides the high-energy gamma-ray emission discussed in this work,
the high-energy neutrino emission arising from $p\gamma$
interactions between shock-accelerated protons and photons from
the supernova may also provide a constraint on the model for
low-luminosity GRBs (e.g. Yu et al. 2008).

Trans-relativistic ejecta may also exist in the usual high
luminosity long GRBs, besides in low luminosity ones, since
{accelerating shocks are} expected to accompany the supernova.
Berger et al. (2003) and Sheth et al. (2003) found that the radio
and optical afterglow indicates a low velocity component more than
$1.5 \rm{days}$ after the explosion in GRB030329/SN2003dh.
However, since the  highly relativistic ejecta is much more
energetic than the trans-relativistic component, high-energy
gamma-ray emission from the latter component {could easily remain
hidden}.

In summary, we have calculated the spectra and light curves of the
high-energy gamma-ray afterglow emission from low luminosity GRBs
for the two main models in the literature, i.e. the
trans-relativistic ejecta model ($\Gamma_0\simeq2$) and the
conventional highly relativistic ejecta model ($\Gamma_0\ga 10$),
considering both synchrotron self inverse-Compton (SSC) and the
external inverse-Compton due to photons from the underlying
supernova/hypernova. Our analysis takes into account a full
Klein-Nishina cross section for inverse Compton scatterings, the
anisotropic scattering of supernova photons and the opacity for
high energy photons due to annihilation with low-energy photons
(mainly from the supernova/hypernova).

We find that for the supernova case the conventional relativistic
outflow model predicts a relatively high gamma-ray flux from SSC
at early times ($<10^5-10^6 {\rm s}$ for typical parameters) with
a rapidly decreasing flux, while in the trans-relativistic outflow
model, a much flatter light curve of high-energy gamma-ray
emission dominated by the SSC emission, is expected at early
times. For the hypernova case, the SSC emission also dominates and
decays sharply at early time in the conventional relativistic
ejecta model, while for the trans-relativistic ejecta model, both
the SSC emission and the SNIC emission could be dominant at early
time, depending on the shock parameters $\epsilon_e$ and
$\epsilon_B$. The main difference between these two models arises
from their different initial Lorentz factors, which induces a
different dynamical evolution of the shock at early times. As a
result, different observational features arise, such as different
light curve shapes, different flux levels and different dominant
components (as detailed in $\S$ \ref{spectra and light curves}).
As shown in $\S$ \ref{detectability of Fermi LAT}, high-energy
gamma-ray emission can be detected in both models as long as
$\epsilon_e$ is large enough, although detection from the
conventional relativistic ejecta  is much easier. Thus, with
future high energy gamma-ray observations by Fermi LAT, one can
expect to be able to distinguish between the two models based on
the above differences in observational features.

{\acknowledgments  We would like to thank Z. G. Dai, Y. F. Huang
and D. M. Wei for useful discussions. This work is supported by
the National Natural Science Foundation of China under grants
10973008 and 10403002, the National Basic Research Program of
China (973 program) under grants No. 2009CB824800, the Foundation
for the Authors of National Excellent Doctoral Dissertations of
China, the Qing Lan Project, and NASA grant NNX 08AL40G.}

\begin{figure}
\epsscale{.80} \plotone{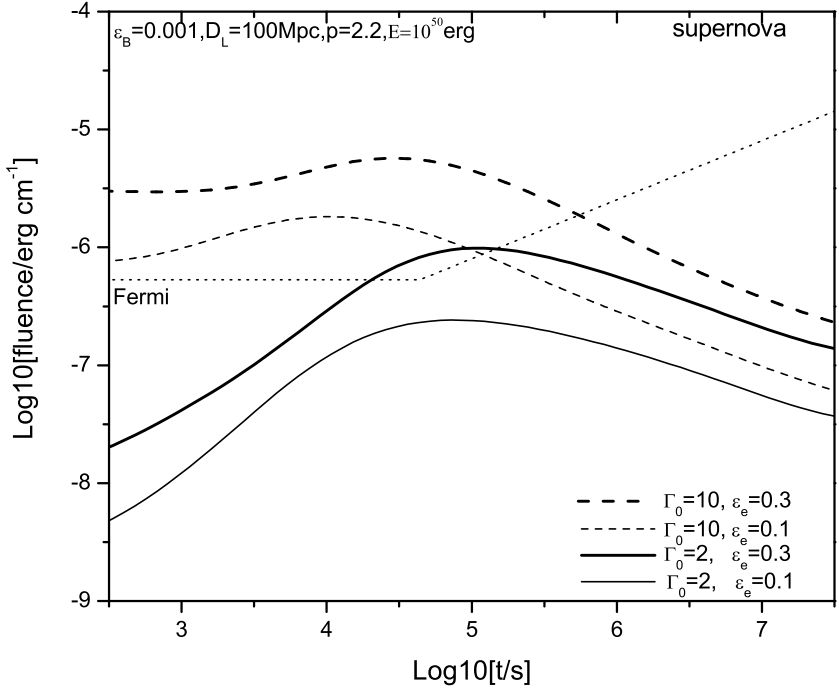} \caption{Time-integrated fluence
(defined as $\int_{0.5t}^{t}\int_{\nu_{1}}^{\nu_{2}}F_{\nu}d\nu
dt$, where $h\nu_{1}=20{\rm MeV}$ and $h\nu_{2}=\min(\nu_{\rm
cut},300{\rm GeV})$) of the high-energy gamma-ray emission in the
two models of low-luminosity GRBs. Short dotted lines represent
the fluence threshold of Fermi LAT at $400{\rm MeV}$ with
effective detection area $A_{\rm eff}=6000{\rm cm^{2}}$. The
thicker and thinner solid lines represent the time-integrated
fluence in the trans-relativistic ejecta model with
$\epsilon_e=0.3$ and $\epsilon_e=0.1$, respectively; the dashed
lines and dotted lines represent the time-integrated fluence in
the conventional relativistic ejecta model with $\epsilon_e=0.3$
and $\epsilon_e=0.1$, respectively. Other parameteres are
$E=10^{50}\rm{erg}$, $\dot{m}=1$ and $\epsilon_B=0.001$.
\label{Fermi_E1}}
\end{figure}

\begin{figure}
\epsscale{.80} \plotone{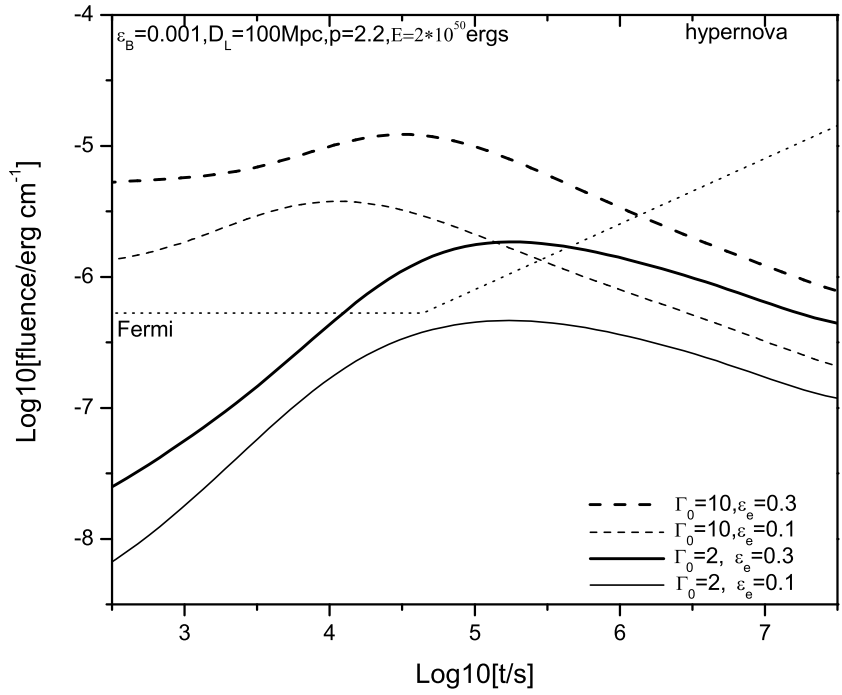} \caption{Same as
Fig.\ref{Fermi_E1}, but for the case of a hypernova explosion that
has a larger total energy ($E=2\times10^{50}\rm{erg}$), in
accordance with the parameters used in Ando \& \Meszaros (2008).
\label{Fermi_E2_hyp}}
\end{figure}


\clearpage
\begin{thebibliography}{99}
\bibitem[Aharonian \& Atoyan(1981)]{1981Ap&SS..79..321A} Aharonian, F.~A., \& Atoyan, A.~M.\ 1981, \apss, 79, 321
\bibitem[Ando \& M{\'e}sz{\'a}ros(2008)]{2008ApJ...689..351A} Ando, S., \& M{\'e}sz{\'a}ros, P.\ 2008, \apj, 689, 351
\bibitem[Atwood et al.(2009)]{2009ApJ...697.1071A} Atwood, W.~B., et al.\ 2009, \apj, 697, 1071
\bibitem[Baring \& Harding(1997)]{1997ApJ...491..663B} Baring, M.~G., \& Harding, A.~K.\ 1997, \apj, 491, 663
\bibitem[Berger et al.(2003)]{2003ApJ...599..408B} Berger, E., Kulkarni, S.~R., Frail, D.~A., \& Soderberg, A.~M.\ 2003, \apj, 599, 408
\bibitem[Blandford \& McKee(1976)]{1976PhFl...19.1130B} Blandford, R.~D., \& McKee, C.~F.\ 1976, Physics of Fluids, 19, 1130
\bibitem[Blumenthal \& Gould(1970)]{1970RvMP...42..237B} Blumenthal, G.~R., \& Gould, R.~J.\ 1970, Reviews of Modern Physics, 42, 237
\bibitem[Brunetti(2000)]{2000APh....13..107B} Brunetti, G.\ 2000, Astroparticle Physics, 13, 107
\bibitem[Campana et al.(2006)]{2006Natur.442.1008C} Campana, S., et al.\ 2006, \nat, 442, 1008
\bibitem[Chevalier \& Fransson(2008)]{2008ApJ...683L.135C} Chevalier, R.~A., \& Fransson, C.\ 2008, \apjl, 683, L135
\bibitem[Chevalier \& Li(1999)]{1999ApJ...520L..29C} Chevalier, R.~A., \& Li, Z.-Y.\ 1999, \apjl, 520, L29
\bibitem[Cobb et al.(2006)]{2006ApJ...645L.113C} Cobb, B.~E., Bailyn, C.~D., van Dokkum, P.~G., \& Natarajan, P.\ 2006, \apjl, 645, L113
\bibitem[Colgate(1974)]{1974ApJ...187..333C} Colgate, S.~A.\ 1974, \apj, 187, 333
\bibitem[Cusumano et al.(2006)]{2006GCN..4786....1C} Cusumano, G., Moretti, A., Tagliaferri, G., Kennea, J., \& Burrows,D.\ 2006, GRB Coordinates Network, 4786, 1
\bibitem{} Dai, X. Y. 2008, ApJ, submitted, arXiv:0812.4466
\bibitem[Fan et al.(2006)]{2006JCAP...09..013F} Fan, Y.-Z., Piran, T., \& Xu, D.\ 2006, Journal of Cosmology and Astro-Particle Physics, 9, 13
\bibitem[Fan et al.(2008)]{2008MNRAS.384.1483F} Fan, Y.-Z., Piran, T.,
Narayan, R., \& Wei, D.-M.\ 2008, \mnras, 384, 1483
\bibitem[Fan \& Piran(2006)]{2006MNRAS.370L..24F} Fan, Y., \& Piran, T.\ 2006, \mnras, 370, L24
\bibitem[Galama et al.(1998)]{1998Natur.395..670G} Galama, T.~J., et al.\ 1998, \nat, 395, 670
\bibitem[Gou \& M{\'e}sz{\'a}ros(2007)]{2007ApJ...668..392G} Gou, L.-J., \& M{\'e}sz{\'a}ros, P.\ 2007, \apj, 668, 392
\bibitem[Guetta et al.(2004)]{2004ApJ...615L..73G} Guetta, D., Perna, R., Stella, L., \& Vietri, M.\ 2004, \apjl, 615, L73
\bibitem{} Guetta, D. \& Della Valle, M.,  2007, ApJ, 657, L73
\bibitem[Heitler(1954)]{1954qtr..book.....H} Heitler, W.\ 1954, International Series of Monographs on Physics, Oxford: Clarendon,
1954, 3rd ed.,
\bibitem[Huang et al.(1999)]{} Huang, Y. F., Dai, Z. G. \& Lu, T.
1999, MNRAS, 309, 513
\bibitem[Huang et al.(2000)]{2000ApJ...543...90H} Huang, Y.~F., Gou, L.~J., Dai, Z.~G., \& Lu, T.\ 2000, \apj, 543, 90
\bibitem{}Huang, Y. F. \& Cheng, K. S., 2003, MNRAS, 341, 263
\bibitem[Iwamoto et al.(1998)]{1998Natur.395..672I} Iwamoto, K., et al.\ 1998, \nat, 395, 672
\bibitem[Kulkarni et al.(1998)]{1998Natur.395..663K} Kulkarni, S.~R., et al.\ 1998, \nat, 395, 663
\bibitem[Liang et al.(2006)]{2006ApJ...653L..81L} Liang, E.-W., Zhang, B.-B., Stamatikos, M., Zhang, B., Norris, J., Gehrels, N., Zhang, J., \& Dai, Z.~G.\ 2006, \apjl, 653, L81
\bibitem[Lithwick \& Sari(2001)]{2001ApJ...555..540L} Lithwick, Y., \& Sari, R.\ 2001, \apj, 555, 540
\bibitem[Matzner \& McKee(1999)]{1999ApJ...510..379M} Matzner, C.~D., \& McKee, C.~F.\ 1999, \apj, 510, 379
\bibitem[Mazzali et al.(2006)]{2006Natur.442.1018M} Mazzali, P.~A., et al.\ 2006, \nat, 442, 1018
\bibitem[Mirabal et al.(2006)]{2006ApJ...643L..99M} Mirabal, N., Halpern, J.~P., An, D., Thorstensen, J.~R.,
\& Terndrup, D.~M.\ 2006, \apjl, 643, L99
\bibitem[Moderski et al.(2000)]{2000ApJ...529..151M} Moderski, R., Sikora,
M., \& Bulik, T.\ 2000, \apj, 529, 151
\bibitem[Nakar(2007)]{2007PhR...442..166N} Nakar, E.\ 2007, \physrep, 442, 166
\bibitem[Panaitescu et al.(1998)]{1998ApJ...503..314P} Panaitescu, A., Meszaros, P., \& Rees, M.~J.\ 1998, \apj, 503, 314
\bibitem[Pian et al.(2006)]{2006Natur.442.1011P} Pian, E., et al.\ 2006, \nat, 442, 1011
\bibitem[Sakamoto et al.(2006)]{2006GCN..4822....1S} Sakamoto, T., et al.\ 2006, GRB Coordinates Network, 4822, 1
\bibitem[Sari \& Esin(2001)]{2001ApJ...548..787S} Sari, R., \& Esin, A.~A.\ 2001, \apj, 548, 787
\bibitem[Sari \& Piran(1995)]{1995ApJ...455L.143S} Sari, R., \& Piran, T.\ 1995, \apjl, 455, L143
\bibitem{} Sazonov, S. Y.; Lutovinov, A. A.; Sunyaev, R. A. 2004;
Nature, 430, 646
\bibitem[Sheth et al.(2003)]{2003ApJ...595L..33S} Sheth, K., Frail, D.~A., White, S., Das, M., Bertoldi, F., Walter, F., Kulkarni, S.~R., \& Berger, E.\ 2003, \apjl, 595, L33
\bibitem[Soderberg et al.(2006)]{2006Natur.442.1014S} Soderberg, A.~M., et al.\ 2006, \nat, 442, 1014
\bibitem[Soderberg et al.(2008)]{2008Natur.454..246S} Soderberg, A.~M., et al.\ 2008, \nat, 454, 246
\bibitem[Stecker et al.(1992)]{1992ApJ...390L..49S} Stecker, F.~W., de Jager, O.~C., \& Salamon, M.~H.\ 1992, \apjl, 390, L49
\bibitem[Tan et al.(2001)]{2001ApJ...551..946T} Tan, J.~C., Matzner, C.~D., \& McKee, C.~F.\ 2001, \apj, 551, 946
\bibitem[Toma et al.(2006)]{2006ApJ...640L.139T} Toma, K., Ioka, K., Yamazaki, R., \& Nakamura, T.\ 2006, \apjl, 640, L139
\bibitem[Wang et al.(2007)]{2007ApJ...664.1026W} Wang, X.-Y., Li, Z., Waxman, E., \& M{\'e}sz{\'a}ros, P.\ 2007, \apj, 664, 1026
\bibitem{} Wang, X.-Y., Li, Z. \& M{\'e}sz{\'a}ros, P.\ 2006, \apj, 641, L89
\bibitem{} Wang, X.-Y. \& M{\'e}sz{\'a}ros, P.\ 2006, \apj, 643, L95
\bibitem[Waxman(2004)]{2004ApJ...602..886W} Waxman, E.\ 2004, \apj, 602, 886
\bibitem[Waxman et al.(2007)]{2007ApJ...667..351W} Waxman, E., M{\'e}sz{\'a}ros, P., \& Campana, S.\ 2007, \apj, 667, 351
\bibitem[Woosley et al.(1999)]{1999ApJ...516..788W} Woosley, S.~E., Eastman, R.~G., \& Schmidt, B.~P.\ 1999, \apj, 516, 788
\bibitem[Yu et al.(2007)]{2007ApJ...671..637Y} Yu, Y.~W., Liu, X.~W., \& Dai, Z.~G.\ 2007, \apj, 671, 637
\bibitem[Yu et al.(2008)]{2008MNRAS...385..1461Y} Yu, Y.~W., Dai, Z.~G., \& Zheng X. P., 2008, \mnras, 385, 1461
\bibitem[Zel{\'d}ovich et al.(2002)]{zeldovich 2002} Zel{\'d}ovich, Ya. B., \& Raizer, Yu. P. 2002,
Physics of Shockwaves and High Temperature Hydrodynamic Phenomena
(Mineola: Dover)
\bibitem[Zhang \& M{\'e}sz{\'a}ros(2001)]{2001ApJ...552L..35Z} Zhang, B., \& M{\'e}sz{\'a}ros, P.\ 2001, \apjl, 552, L35


\end {thebibliography}
\end{document}